\def\sumk{\int_{\rm{BZ}}\ \frac{d^3\k}{8\pi^3}\ }
\def\uu#1{\underline{\underline{#1}}}

\def\ket{\vert \vert  \{ \emptyset \} \rangle}
\def\ket2{\vert \vert \otimes \{ R \} \rangle}

\def\r{{\mathbf{r}}_i}
\def\n{\noindent }
\def\mbf#1{{\mathbf {#1}}}
\def\jg{{\mbf j}_\gamma}
\def\wt#1{\widetilde{\mbf{#1}}}
\def\eq{\ =\ }
\def\mns{\ -\ }
\def\pls{\ +\ }
\def\be{\begin{equation}}
\def\ee{\end{equation}}
\def\<{[}
\def\>{]}

\def\k{{\mathbf k}}

\def\G{ G}

\def\f{\mbf f}
\def\j{\mbf j}
\documentclass[twocolumn,showpacs,preprintnumbers,amsmath,amssymb]{revtex4}

\usepackage{graphicx}
\usepackage{dcolumn}
\usepackage{bm}
\begin{document}

\title{Optical properties of random alloys : A formulation}

\author{Kamal Krishna Saha}
\email{kamal@bose.res.in}
\author{Abhijit Mookerjee}
\affiliation{ S. N. Bose National Centre for Basic Sciences. Block-JD, Sector-III, \\
Salt Lake City, Kolkata-700098, India.}

\begin{abstract}
{We present here a formulation for the calculation of the configuration-averaged optical conductivity in 
random alloys. Our formulation is based on the augmented-space theorem introduced by one of us 
[A. Mookerjee, J. Phys. C: Solid State Phys. {\bf 6}, 1340 (1973)]. We show that disorder 
scattering renormalizes the electron and hole propagators
as well as the transition amplitude. The corrections to the transition amplitude have
been shown to be related to the self-energy of the propagators and vertex corrections.} 
\end{abstract}
\date{\today}

\pacs{71.23.-k}
\maketitle
\parindent 0pt

\section{Introduction}

The object of our present study is to present a formulation for obtaining  the configuration-averaged optical
conductivity for random alloys. Because of randomness, there is a  need  to go beyond the
usual reciprocal-space-based formulations for crystalline compounds.  
Instead of labelling the electronic states by 
the Bloch wave vector and band index ($\k,j$), which is   suitable for
crystalline compounds, we have to label them   
 by energy and the  composite angular momentum  $L$ = ($\ell,m,m_s$).
In cases where the disorder is substitutional and {\sl homogeneous}, in the sense that the occupation 
probabilities of lattice sites by atom species are independent of the site label, we can still label the
configuration-averaged quantities by the reciprocal wave vectors. However, the band picture breaks down, and
we cannot use the band index labeling of quantum states as in crystalline materials.
Substitutional disorder dictates that we begin with  a purely real-space representation and we have chosen 
 the minimal basis set of the tight-binding linearized muffin-tin orbitals (TB-LMTO) method \cite{Ander1,Ander2}.
Configuration averaging over various random atomic arrangements has been carried out 
using the augmented-space  formalism   (ASF) introduced by us earlier for the study of electronic properties of
disordered systems \cite{Am,Am3}. The ASF goes beyond the usual
mean-field approaches and takes into account configuration fluctuations. This formalism has been described in detail 
in the referenced articles and the interested reader can go into the details in them.\linebreak            
The contribution of the paper will be to show that the disorder-induced corrections to
the averaged current terms in the optical conductivity are directly related either to
the disorder-induced self-energy in the propagators or to vertex corrections. Since
the self-energy and the vertex corrections can be calculated for realistic binary
alloy systems, either within an augmented-space recursion \cite{Am2,Am4,Am3} or 
within one of the mean-field approaches \cite{KG}, this formulation will form the
basis of subsequent calculations in real alloys.

\section{The optical conductivity}

The Hamiltonian describing the effect of an external  radiation field on the electronic states of a solid is
given by

\[ H\eq  \sum_{i=1}^{N}\left\{ \frac{1}{2}\left[ {\mathbf{p}}_i + {\mathbf{A}} ({\mathbf{r}}_i,t)\right]^2 
+ V({\mathbf{r}}_i) + \phi(\r,t)\right\}.
\]

Here ${\mathbf{A}}(\r,t)$ and $\phi(\r,t)$ are the vector and scalar potentials seen by the $i$th electron because of the
radiation field.
We have used atomic units in which the electronic charge, mass, the Planck constant, and the velocity of
light are set to unity ($e$=1, $m$=1, $\hbar$=1, $c$=1). There are $N$ electrons whose positions are labelled by $\r$.
The effective potential $V({\bf r}_i)$ experienced by individual  electrons is expressed 
in the local-density approximation (LDA) of the density-functional theory (DFT). 
To first order in the vector potential, the Hamiltonian becomes

\be
H\eq  \sum_{i=1}^{N}\left\{ \frac{1}{2}{\mathbf{p}}_i^2 \pls  V({\mathbf{r}}_i) \pls {\mathbf {A}}({\mathbf {r}}_i,t)\cdot {\mathbf {j}}_i \right\}.
\ee

The current operator {\bf j}$_i$ is related to the momentum operator as $(e/m)\ {\bf p}_i$. In atomic
units, {\bf j}$_i$ = {\bf p}$_i$ = {\bf v}$_i$ = $d \r/dt$. We work in the Coulomb gauge where  
${\mathbf {\nabla}}\cdot {\mathbf{A}}(\r,t)$ = 0 and $\phi(\r,t)$ = 0, so that the electric field

\[ {\mathbf{E}}(\r,t) =  - \frac{\partial {\mathbf{A}}(\r,t)}{\partial t}.  \]

The Kubo formula relates the linear current response to the radiation field,

\[ \langle j_{\mu}(t) \rangle \eq \sum_{\nu}\int_{-\infty}^{\infty} \ dt'\ \chi_{\mu\nu}(t-t') A_{\nu} (t'), \]

\[ \chi_{\mu\nu}(\tau) \eq  \imath \Theta(\tau)\ \langle \Phi_0\vert [j_{\mu}(\tau),j_{\nu}(0)]\vert \Phi_0\rangle, \]

where $\tau = t-t'$ and $\Theta(\tau)$ is the Heaviside step function,

\[ \Theta(\tau) \eq \left\{ \begin{array}{ll}
                             1 & \mbox{if } \tau\ >\ 0 \\                               
                             0 & \mbox{if } \tau \ \leq \ 0 .
                            \end{array} \right. \]

$\vert\Phi_0\rangle$ is the ground state of the unperturbed system, that is, the electrons in the solid in the absence of the radiation field. In
 the absence of the radiation field, there is no photocurrent, i.e., $\langle\Phi_0\vert j_{\mu}\vert\Phi_0\rangle$ = $0$.
The fluctuation-dissipation theorem  relates the imaginary part of the generalized susceptibility to the
correlation function as follows:

\be
\chi_{\mu\nu}^{\prime\prime}(\omega) \eq \frac{1}{2} \left(1-e^{-\beta\omega}\right) S_{\mu\nu}(\omega),
\ee

\noindent where $ \beta = 1/{k_B T}$; $k_B$ is the Boltzmann constant, $T$ the temperature, and  

\[ \chi_{\mu\nu}^{\prime\prime}(\omega)\eq \mbox{Im} \int_{-\infty}^{\infty}\ dt\ e^{\imath z\tau}\ \chi_{\mu\nu}(\tau), \quad \mbox{z=$\omega$+$\imath 0^+$} \]

\noindent and

\[S_{\mu\nu}(\omega) \eq \mbox{Im} \int_{-\infty}^{\infty}\ dt\ e^{\iota z\tau}\ \langle\Phi_0\vert j_{\mu}(\tau)\ 
j_{\nu}(0)\vert \Phi_0\rangle, \quad \mbox{z=$\omega$+$\imath 0^+$}. \]

An expression for the correlation function, at $T=0$ K,  can be obtained via the Kubo-Greenwood expression,

\begin{eqnarray}
S(\omega)\eq \frac{1}{3\pi}\ \sum_{\gamma}\ \int dE\ \mbox{Tr}\ \left[\rule{0mm}{4mm}\ {\mathbf j}_{\gamma}\ 
\mbox{Im} \{ {\mathbf G}^v ( E)\}\ {\mathbf j}_{\gamma}^\dagger\ \right. \nonumber \\
\left.\mbox{Im} \{{\mathbf G}^c (E+\omega)\} \right]. \quad
\label{ref1}
\end{eqnarray}

We have assumed isotropy of the response so that the tensor $S_{\mu\nu}$ is diagonal and we have defined $S(\omega)$ as 
the direction averaged quantity $\frac{1}{3}\sum_{\mu} S_{\mu\mu}(\omega)$. 
$\jg$ is $\hat{e}_\gamma\cdot{\bf j},$ and  $\hat{e}_{\gamma}$ is the direction of polarization of the incoming photon.
The operators ${\bf G}^v(E)$ and ${\bf G}^{c}(E)$ are the resolvents of the Hamiltonian projected 
onto the subspaces spanned by the occupied and the unoccupied one-electron states. 

The trace is invariant in different representations. For crystalline systems, usually the Bloch basis 
$\{\vert {\mathbf k},j\rangle\}$ is used. For disordered systems, prior to configuration
averaging, it is more convenient to use the real-space
based screened (or tight-binding) muffin-tin orbitals as a basis $\{\vert RL\rangle\}$. 

\noindent If we define
\be 
{\cal S}_\gamma(z_1,z_2)  = \mbox{Tr}\ \left[\ \rule{0mm}{4mm}{\mathbf j}_{\gamma}\ {\mathbf G}^v (z_1)\ 
{\mathbf j}_{\gamma}^\dagger\ {\mathbf G}^c (z_2)\right],
\label{eq4}
\ee

\noindent  then the above equation becomes

\begin{eqnarray}
S(\omega) &\eq& \frac{1}{12\pi}\sum_{\gamma}\ \int dE\ \left[\ {\cal S}_\gamma(E^-,E^{+}+\omega) 
 \ldots\right. \nonumber\\
&&\left.\dots + {\cal S}_\gamma(E^+,E^{-}+\omega)-{\cal S}_\gamma(E^+,E^{+}+\omega)  \right.\dots \nonumber\\
&&\left.\dots -{\cal S}_\gamma(E^-,E^{-}+\omega)\right.],
\end{eqnarray}

\noindent where

\[ f(E^\pm) \eq \lim_{\delta\rightarrow 0} f(E\pm i\delta). \]

We have used the Herglotz property of the Green operator,

\[{\mbf G}(E+i\delta) = {\mbf G}^{r}(E)\mns  i\ \mbox{sgn}(\delta)\ {\mbf G}^i(E). \]

For disordered materials, we shall be interested in obtaining the configuration averaged  response functions. This will require
the configuration averaging of quantities such as ${\cal S}_\gamma(z_1,z_2)$.

\section{Configuration averaging}

Any description of a disordered system must be from a statistical point of view, since the properties of these systems are random
variables and any particular {\sl configuration} is of little interest \footnote{This statement has to be 
modified in some situations, like localized states in band tails, where unlikely configurations play an
important role and configuration averaging is not meaningful.}. Consequently, the study of 
configuration-averaged properties
of disordered systems has received much attention.

Configuration averaging for response functions in disordered materials has had some history. The Ziman-Faber theory 
\cite{ziman,bradley,fz}, much in use for liquids, is valid for electrons, weakly scattered from a dilute distribution of
impurities. The extended version of this theory was proposed by Evans {\em et al.} \cite{evans}, but this too overlooks 
multiple scattering effects, as pointed out by Roth and Singh \cite{rs}. An effective medium approach (EMA) was proposed by 
Roth \cite{roth} and developed further by Roth and Singh \cite{rs,roth,rs1} and Asano and Yonezawa \cite{ay}. 
This approach does take into account multiple scattering effectively and, as we
shall see, will have close similarities with the approach we propose in this paper. Velick\'y \cite{v} has developed an expression
for configuration-averaged response functions in random alloys for simple tight-binding models with one orbital per
site and diagonal disordered within the coherent-potential approximation (CPA).
Brouers and Vedyayev \cite{bv} have extended the formalism to transition-noble metal alloys. The CPA-like mean-field approach has been
applied to response functions by Niizeki and co-workers \cite{nii,hw,nh}, who extended the pioneering work of Velick\'y 
\cite{v} to longer-ranged random potentials. Mookerjee \cite{Am} has introduced the ASF
 to tackle configuration averaging. Within this formalism he studied the role of macroscopic conservation
laws on the response functions, leading to a Ward identity between the vertex corrections and the self-energy \cite{mook2}.
Within the CPA, vertex corrections were obtained by ingenious diagram summations by Leath \cite{leath}. There have been
CPA calculations by Harris and Plischke \cite{hp} and Nauciel-Bloch and Riedinger \cite{nr}. In a series of papers, Mookerjee and 
co-workers \cite{mook3,mty,mt} have applied the ASF to conductivity and optical conductivity in random
alloys. This will form the background of our present development.

The ASF has been described thoroughly in a series of articles \cite{Am,KG,Am2,Am4,Am3}.
 Here we shall introduce the salient features which will be required by us
subsequently in this paper.  We shall start from a 
first-principles tight-binding linearized muffin-tin orbitals (TB-LMTO) \cite{Ander1,Ander2} method. 

The quenched local randomness in the alloy is described by a set  of random {\sl occupation} variables $n_{R }$,
which takes the value 1 if the muffin-tin labelled by $\{R \}$ is occupied by an $A$-type atom and 0 if it is 
occupied by a $B $. The atom sitting at $\{R \}$ can either be of the type $A$ ($n_{R}=1$) with probability 
$x_A$ or $B$ ($n_{R}=0$) with probability $x_B$.                                                             

The ASF now introduces the configuration space $\Phi$ of the set of $N$ random variables $\{n_{R }\}$ 
of rank $2^N$ spanned
by configurations of the kind $\vert \uparrow\uparrow\downarrow\cdots\downarrow\uparrow\cdots\rangle$,  where                

\begin{eqnarray*}                                                                           
\vert\uparrow_{R }\rangle &\eq& \sqrt{x_A} \ \vert {A_{ }}_R \rangle \pls \sqrt{x_B} \ \vert {B_{ }}_R \rangle \\
\vert\downarrow_{R }\rangle &\eq& \sqrt{x_B} \ \vert {A_{ }}_R \rangle \mns \sqrt{x_A} \ \vert {B_{ }}_R \rangle.
\end{eqnarray*}

If we define the
configuration $\vert\uparrow\uparrow\cdots\uparrow\cdots\rangle$ as the {\sl average} or {\sl reference} configuration,
then any other configuration may be uniquely labelled by the {\sl cardinality sequence}, $\{ R_k\}$,  which is the sequence
  of
positions where we have a $\downarrow$ configuration. The {\sl cardinality sequence} of the {\sl reference} configuration
is the null sequence $\{\emptyset\}$.

The augmented-space theorem \cite{Am} states that

\be 
\ll A(\{ n_{R }\}) \gg \eq < \{\emptyset\}\vert \widetilde{{\mbf A}}\vert \{\emptyset\}>,
\ee

\n where

\[\widetilde{\mbf A}(\{{\bf M}_R\}) \eq \int \cdots \int A(\{\lambda_{R }\})\ \prod d{\mbf P}(\lambda_{R }) 
\quad \epsilon\quad \Phi . \]

\n {\bf P}($\lambda_{R }$) is the spectral density of the self-adjoint operator ${{\mbf M}}_{R }$, which is
such that the probability density of $n_{R }$ is given by 

\[ p(n_{R })\eq -\frac{1}{\pi}\ \lim_{\delta\rightarrow 0}\ \mbox{Im} \ \langle \uparrow\vert 
\left(\rule{0mm}{4mm}(n_{R }+i\delta) {{\mbf I}}\mns {{\mbf M}}_{R }\right)^{-1}\vert\uparrow\rangle .\]

\n  ${{\bf M}}_{R }$ is an operator in the space of configurations $\psi_{R }$ of the variable $n_{R }$. 
This is of rank 2 and is spanned by the {\sl states} $ \{ \vert \uparrow _{R } \rangle, \vert \downarrow _{R } 
\rangle \} $,

\begin{equation}
{{\mathbf M}}_{R } \ = \ x_A \ {\cal P}^{\uparrow}_{R } \ + \ x_B \ {\cal P}^{\downarrow}_{R } \ 
+\ \sqrt{x_A x_B }\ \left( {\cal T}^{\uparrow \downarrow}_{R } + {\cal T}^{\downarrow\uparrow}_{R } \right),
\end{equation}

\n where ${\cal P}^{\uparrow}_{R }=\vert\uparrow_R\rangle\langle\uparrow_R\vert$,
${\cal P}^{\downarrow}_{R }=\vert\downarrow_R\rangle\langle\downarrow_R\vert$, and
${\cal T}^{\uparrow \downarrow}_{R }=\vert\uparrow_R\rangle\langle\downarrow_R\vert$
are projection and transfer operators in configuration-space. 

\n Within the ASF, the configuration-averaged Green function is given by \cite{gdma}~:

\begin{equation}
\ll \mbf{G}(z) \gg  \eq  \langle 1 \vert \left( z\mbf{\widetilde{I}}\mns 
\mbf{\widetilde{H}}^{\rm{eff}}\right)^{-1} \vert 1\rangle, 
\label{avg}
\end{equation}

\begin{widetext} 
\begin{eqnarray}
\mbf{\widetilde{H}}^{\rm{eff}}  =  \sum_{R}\left\{\rule{0mm}{5mm} \widetilde{\mathbf A}\ 
{\cal P}_R \otimes\widetilde{\cal I} + \widetilde{\mathbf B}\ {\cal P}_{R} \otimes {\cal P}^{\downarrow}_{R} +
\widetilde{\mathbf F}\ {\cal P}_{R}  \otimes \left( {\cal T}_{R}^{\uparrow\downarrow} + {\cal
 T}_{R}^{\downarrow\uparrow} \right) +\sum_{R'}\ \widetilde{\mathbf V}_{RR'}\ {\cal T}_{RR'} \otimes 
{\mathaccent "7E {\cal I}}\right\}, \nonumber\\
\label{kasr2}
\end{eqnarray}
\n where
\begin{eqnarray}
&&\widetilde{\mathbf A}\ \equiv\ \widetilde{A}_{L }\ \delta_{LL'}  \eq   A\left( C_{L }/\Delta_{L } \right) / A\left(
     1/\Delta_{L } \right)\ \delta_{LL'}, \nonumber\\
&&\widetilde{\mathbf B}\ \equiv\ \widetilde{B}_{L }\ \delta_{LL'}  \eq   B\left[ (C_{L }-z)/\Delta_{L }\right] / A\left( 1/\Delta_{L }
     \right) \ \delta_{LL'}, \nonumber\\
&&\widetilde{\mathbf F}\ \equiv\ \widetilde{F}_{L }\ \delta_{LL'} \eq  F\left[ (C_{L }-z)/\Delta_{L }\right] / A\left( 1/\Delta_{L }
      \right)\ \delta_{LL'},  \nonumber\\
&&\widetilde{\mathbf V}\ \equiv\  \widetilde{V}_{LL'}(R-R') \eq  A\left( 1/\Delta_{L }
     \right)^{-1/2}\ S_{LL'}(R-R') \ A\left( 1/\Delta_{L'} \right)^{-1/2}.
\end{eqnarray}
\end{widetext}

\n ${\cal P}_R=\vert R\rangle\langle R\vert$ and ${\cal T}_{RR'}=
\vert R\rangle\langle R'\vert$ are projection and transfer operators in real-space, and $L$ is a composite angular momentum index 
$\{\ell,m,m_s\}$.  $C, \Delta$ are the TB-LMTO potential parameters and $S$ is the structure matrix,  in the most
tight-binding ($\alpha$) representation \cite{Ander1,Ander2}. The following functions are~:  

\begin{eqnarray*}
A(y ) & = &  x_A\  y_{A } + x_B\ y_{B }, \quad \mbox{i.e.}, \mbox{ the average of }y,  \\
B(y ) & = &  (x_B - x_A ) (y_{A } - y_{B }), \\
F(y ) & = & \sqrt{x_B x_A} (y_{A } - y_{B }),
\end{eqnarray*}

\n and

\[\vert \mbox{\rm 1} \} = A\left(\Delta_{L }^{-1/2} \right) \vert \{ \emptyset \} \rangle 
+ F\left( \Delta_{L }^{-1/2} \right)  \vert \{ R  \}\rangle, \quad  \vert {\mathrm 1}\rangle 
= \frac{\vert {\mathrm 1}\}}{\parallel \vert {\mathrm 1}\}\parallel}. \]

We can reformulate the above in a second quantized formalism. This follows the ideas put forward by  Schultz and
Shapero \cite{shultz}, which were extended in the ASF by Mookerjee \cite{mook3}.
For the real-space part, this is straightforward, with a {\sl vacuum}
state described as that one which contains no electron-like excitations, and the fermion creation and annihilation operators are $a^\dagger_{RL }$ and $a_{RL }$ for electrons at the site $R $ with angular momentum indeces $L$. For the configuration-space part, we shall follow the ideas of Ref.~\cite{shultz} and consider the {\sl reference}
state to be the vacuum. Each {\sl spin flip} \footnote{These spins denote configurations rather than electron
spins.} at any site from up to down is then a creation of a configuration fluctuation. Since
each site can have only two configurations, two up to down {\sl spin flips} cannot take place at a site. These excitations are
then {\sl local} and {\sl fermion-like}. Each spin flip from down to up is a destruction of such a {\sl local pseudo-fermion}. The
Fock space is then spanned by all configuration states labelled by the cardinality sequences. The corresponding fermion-like
creation and annihilation operators are $b^\dagger_{R }$ and $b_{R }$. These create and annihilate configuration
fluctuations over the reference state. We should note that the configuration fluctuations are local and quenched.
In second quantized form, the Hamiltonian becomes 

\begin{eqnarray}
&&\widetilde{{\mbf H}} \eq \widetilde{{\mbf H}}_0 \pls \widetilde{{\mbf H}}_1, \nonumber \\
&&\widetilde{{\mbf H}}_0 \eq \sum_{RL }\ \widetilde{A}_{L }\ a^\dagger_{RL } a_{RL } \pls \sum_{RL }
\sum_{R'L'}\ \widetilde{V}_{RL ,R'L'}\ a^\dagger_{RL } a_{R'L'}, \nonumber\\
&&\widetilde{{\mbf H}}_1 \eq \sum_{RL }\left\{\rule{0mm}{4mm} \widetilde{B}_{L }\ a^\dagger_{RL } a_{RL }
\ b^\dagger_{R } b_{R } \right. \nonumber  \\
&&\phantom{xxxxxxxxxxxx}\left. \pls \widetilde{F}_{L }\ a^\dagger_{RL } a_{RL }\ (b_{R }\pls
b^{\dagger}_{R })\right\},
\label{eqh1}
\end{eqnarray}
\n where
\begin{eqnarray*}
&&\left\{ b_{R }, b^\dagger_{R'}\right\} \eq \delta_{RR'}, \\
&&\left\{ b_{R }, b_{R'}\right\} \eq 0\eq \left\{ b^\dagger_{R },b^\dagger_{R'}\right\}, \\
\end{eqnarray*}

\begin{figure}[b]
\begin{center}
\includegraphics[width=3.2in, height=2.0in]{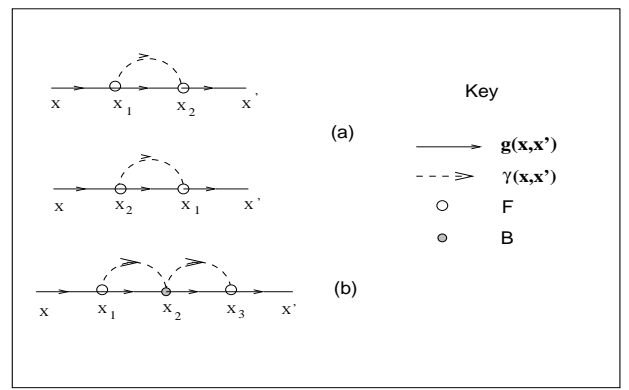}
\caption{\label {fig1}
The scattering diagrams for  (a) the two topologically identical diagrams for $n=2$ and (b) one of the 3! 
topologically identical diagrams for $n=3$.}
\end{center}
\end{figure}

\n and the contraction

\[ b(x')^{\mbf .}\ b(x)^{\dagger\mbf .} \eq i\theta(t'-t)\ \delta_{RR'}, \]

\n where
\[ b(x)\eq b_{R }(t)\eq \exp\left(it\widetilde{{\mbf H}}_0\right)b_{R } \exp\left(-it\widetilde{{\mbf H}}_0\right) \]

\n so that

\begin{eqnarray} 
\gamma(x,x') &\eq& -i\ \langle \{\emptyset\}\vert \left\{ T\ b(x')b^\dagger(x)\right\}\vert\{\emptyset\}\rangle \nonumber \\
&\eq& -i \theta(t'-t)\ \delta_{RR'} \delta(t-t').
\end{eqnarray}

This {\sl pseudo-fermion} formalism has been described earlier by Mookerjee \cite{mook3}. The readers are referred to that
article for further details.

\section{Averaged Green function in the pseudo-fermion formalism}

In this section, we shall develop a multiple scattering formalism for the configuration-averaged Green function
for a random binary alloy. The scattering is by configuration fluctuations and within the second-quantized formalism
just described, the scattering diagrams are Feynman diagrams. The formalism is very close to the Yonezawa-Matsubara
scattering diagrams \cite{ay} and one can establish a one-to-one correspondence between them in the special case of
diagonal disorder.

The augmented-space theorem then states that

\begin{widetext}
\[\ll \mbf{G}(x,x')\gg \eq -i\ \sum_{n=0}^{\infty}\frac{(-i)^n}{n!} \int \ldots\int dt_1dt_2\ldots dt_n \frac{\langle 0\vert \left\{ T\widetilde{{\mbf H}}_1(t_1)\ldots\widetilde{{\mbf H}}_n(t_n)\ a(x)a^{\dagger}(x')\right\}\vert 0\rangle}{\langle 0\vert \widetilde{{\mbf S}}\vert 0\rangle}, \]
\end{widetext}

\n where it is understood that the boldface operators are expressed
by the matrix representation  in $\{ L\}$ space and

\begin{eqnarray*}
&&a(x)\eq a_{RL}(t)\eq \exp\left(it\widetilde{{\mbf H}}_0\right)a_{RL} \exp\left(-it\widetilde{{\mbf H}}_0\right), \\
&&\rule{0mm}{6mm}\widetilde{{\mbf S}} \eq \widetilde{{\mbf U}}(\infty,-\infty) \qquad {\mathrm{and}} \qquad \vert 0\rangle\ 
=\ \vert 0\otimes \{\emptyset\}\rangle, 
\end{eqnarray*}
\n and
\[\widetilde{{\mbf U}}(t,t') \eq \widetilde{{\mbf I}} - \int_{-\infty}^{t} dt''\ \widetilde{{\mbf H}}_1(t'')\ \widetilde{{\mbf U}}(t'',t').\]

We may now apply Wick's theorem  and Feynman's rules and generate a diagrammatic expansion for the averaged Green
function $\ll \mbf{G}(x,x')\gg$ in terms of the VCA Green function, 

\[ {\mbf g} (x,x') \eq -i\ \langle 0\vert \left\{ T\ a(x) a^\dagger(x')\right\}\vert 0\rangle. \]

Let us examine a few terms in the series.

\begin{figure}
\begin{center}
\rotatebox{0}{\includegraphics[width=3.2in, height=2.0in]{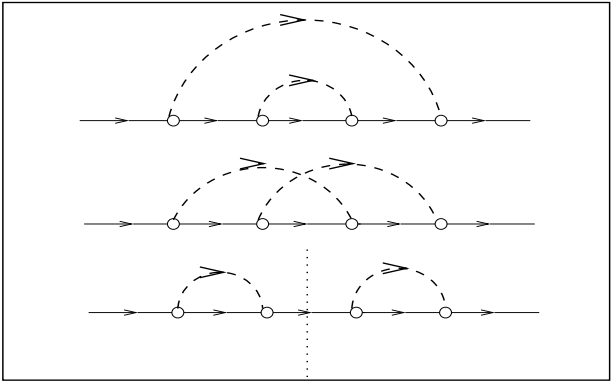}}
\caption{\label {fig2}
The topologically distinct scattering diagrams for $n=4$. Top and middle are non-separable,
the bottom is separable. The middle diagram is a skeleton one.}
\end{center}
\end{figure}

\begin{figure}
\begin{center}
\includegraphics[width=3.2in, height=1.6in]{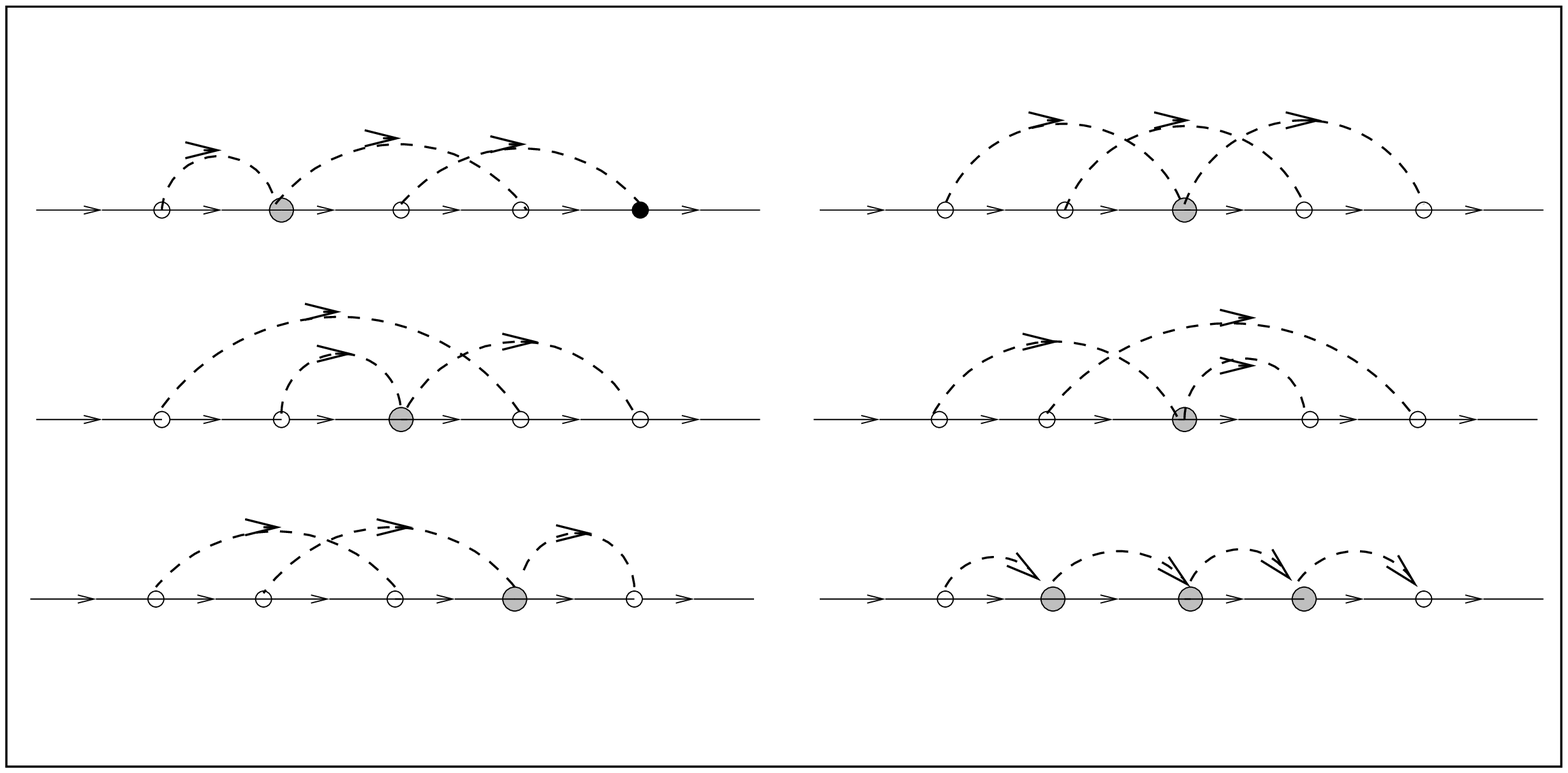}
\caption{\label {fig3}
Skeleton diagrams for order $n=5$.}
\end{center}
\end{figure}

\begin{figure}[t]
\begin{center}
\includegraphics[width=3.0in, height=1.5in]{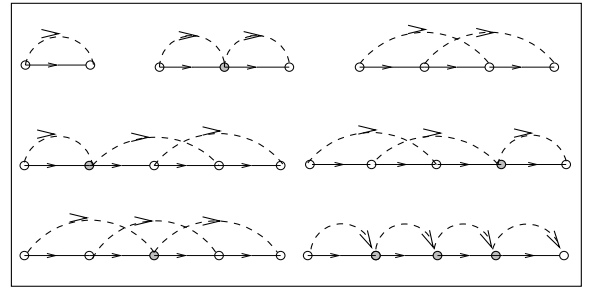}
\caption{\label {fig4}
The skeleton diagrams for the self energy ${\mathbf \Sigma}$(z).}
\end{center}
\end{figure}

\begin{enumerate}

\item[(i)] For $n=1$, the term in the expansion is

\[ -i \int dt_1\ \langle 0\vert \left\{ T\widetilde{{\mbf H}}_1(t_1)a(x)a^\dagger(x')\right\}\vert 0\rangle_{\rm{conn}}.\]

The contribution of this term is zero, since all three terms arising out
of $\widetilde{\bf H}_1$ [see eq. (\ref{eqh1})] vanish because  
 $\langle\{\emptyset\}\vert b^\dagger(x)b(x)\vert\{\emptyset\}\rangle$,
$\langle\{\emptyset\}\vert b^\dagger(x)\vert\{\emptyset\}\rangle$, 
and  $\langle\{\emptyset\}\vert b(x)\vert\{\emptyset\}\rangle$ are all zero.

\item[(ii)] For $n=2$, the only nonvanishing terms come from

\begin{eqnarray*}
&& i^2\ \int dt_1\int dt_2 \langle 0\vert \left\{ T\ {\wt H}_1(t_1){\wt H}_1(t_2)a(x)a^\dagger(x')\right\}\vert 0\rangle_{\rm{conn}} \\
&& = F^2\int dx_1\int dx_2\ \langle 0\vert \left\{ T a^\dagger(x_1)a(x_1) \right.\dots \\
&& \phantom{xx}\left.\dots a^\dagger(x_2)a(x_2)a(x)a^\dagger(x')\right\}\vert 0\rangle_{\rm{conn}} \ldots \\
&& \phantom{xx}\ldots \langle\{\emptyset\}\vert T\ \left( b^\dagger(x_1)+b(x_1)\right)\left(b^\dagger(x_2)+b(x_2)\right)
\vert\{\emptyset\}\rangle_{\rm{conn}} \\
&& \eq F^2\ \left[ \rule{0mm}{4mm} \mbf{g}(x,x_1)\ \mbf{g}(x_1,x_2)\ \mbf{g}(x_2,x')\ \gamma(x_1,x_2)\right.\dots \\
&& \phantom{xxxx}\left.\dots \pls \mbf{g}(x,x_2)\ \mbf{g}(x_2,x_1)\ \mbf{g}(x_1,x')\ \gamma(x_2,x_1)\right]. \\
\end{eqnarray*}

Figure 1 (a) shows a pictorial representation of the two terms, which are topologically identical and therefore have
identical contributions. This cancels the (1/2!) term in the expansion for $\ll \mbf{G}(x,x')\gg$. The $F$ vertex [see eq. (\ref{eqh1})] 
has a contribution $F_{LL'}$ which is
diagonal in $L$ space  $F_{LL'}\ =\ F_L\ \delta_{LL'}$,  where 

\[ F_L=\sqrt{x_Bx_A}\ \left\langle \frac{1}{\Delta_L}\right\rangle^{-1}
\left[\frac{C^A_L}{\Delta_L^A}-\frac{C_L^B}{\Delta_L^B}-z\left(\frac{1}{\Delta_L^A}
-\frac{1}{\Delta_L^B}\right)\right]. \]

\item[(iii)] Figure 1 (b) shows one of the topologically identical diagrams (there are 3! = 6 such diagrams)  for $n=3$. 
Note that it involves the scattering vertex $B$. This
arises from the first term in the expression for $\widetilde{\bf H}_1$ in
eq. (\ref{eqh1}). Its contribution is also diagonal in $L$ space  $B_{LL'}\ =\ B_L\ \delta_{LL'}$, where  

\[  B_L \ =\ \frac{(x_B-x_A)}{\sqrt{x_Bx_A}}\ F_L. \]

 This scattering vertex cannot sit either in the leftmost or  in
the rightmost positions, because one of the associated	pseudo-fermion propagators vanishes.

\item[(iv)] For $n=4$, there are two topologically distinct nonseparable diagrams \footnote{A non-separable
diagram cannot be broken into two along a electron line without also breaking a pseudo-fermion line.} : the double tent and the crossed
tent diagrams shown as the two top diagrams in Fig.~\ref{fig2}. The inner tent in the top diagram goes on to renormalize the interior
Green function from $\mbf{g}(x,x')$ to $\ll \mbf{G}(x,x')\gg$. As such, the middle diagram is the only {\sl skeleton}
diagram at this order. There is one separable diagram (the bottom diagram in Fig.~\ref{fig2}). This can be broken
into two, as shown, by the dotted line. The rightmost diagram renormalizes the rightmost electron line.

\item[(v)] The non-separable topologically distinct diagrams for $n=5$ are shown in Fig.~\ref{fig3}. We note that all
odd-order diagrams must have an odd number of $B$ vertices.
\end{enumerate}

\begin{figure*}[t]
\begin{center}
\framebox{\includegraphics[width=4.2in, height=3.2in]{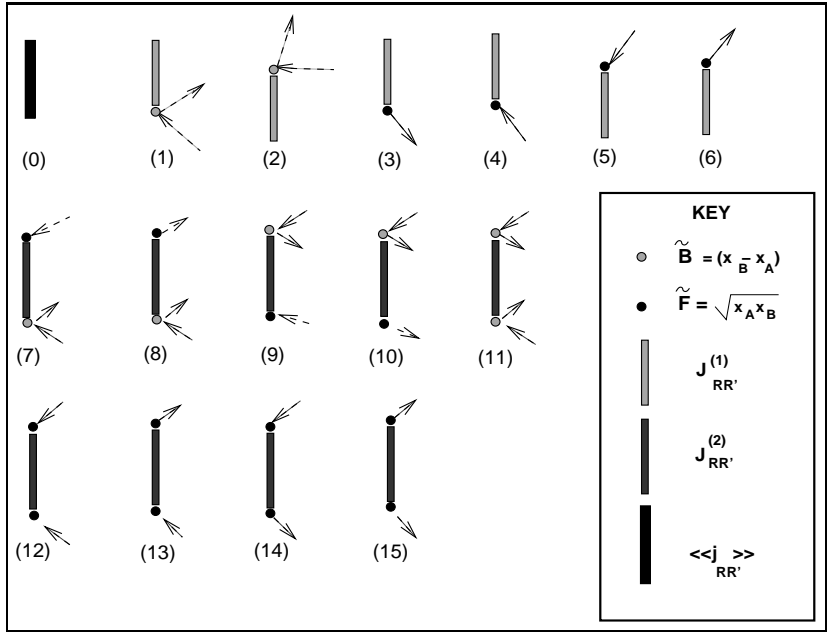}}
\caption{\label {fig5}
The scattering vertices associated with the random current terms.}
\end{center}
\end{figure*}

The skeleton diagrams provide the expression for the self-energy for the Dyson equation which follows~:

\begin{eqnarray*}
&&\ll \mbf{G}(x,x')\gg \eq {\mbf g}(x,x') \dots \\
&& \dots +  \int dy\int dy'\  \mbf{g}(x,y)\ {\mathbf \Sigma}(y,y') \ll \mbf{G}(y',x')\gg .
\end{eqnarray*} 

For homogeneous disorder, we have shown earlier that we have translational symmetry in the full augmented-space \cite{gdma}. 
We can then take the Fourier transform of the above equation to get 

\be  
\mbf{G}({\mbf k},E) \eq \mbf{g}({\mbf k},E) + \mbf{g}({\mbf k},E)\ \mbf{\Sigma}({\mbf k},E)\ \mbf{G}({\mbf k},E) . 
\label{dys}
\ee

The diagrams for the self-energy are shown in Fig.~\ref{fig4}. In the above equation, each term is a matrix
in $\{L\}$ space.

\section{Averaged Optical conductivity in the pseudo-fermion formalism}

We now go back to eq. (\ref{eq4}) and discuss the configuration averaging of the two-particle
Green functions of the kind ${\cal S}_\gamma(z_1,z_2)$. The augmented-space theorem immediately
implies  that
\begin{eqnarray}
&&\ll {\cal S}_\gamma(z_1,z_2)\gg \nonumber \\ 
&&\ \eq \mbox{Tr}\ \left. \left\langle  \{\emptyset\} \left | \
\left[\ \rule{0mm}{4mm} {\wt j}_\gamma\ {\wt G}^v(z_1)\ {\wt j}_\gamma^\dagger\  {\wt G}^c(z_2)\right]\ \right |
\{\emptyset\}\right\rangle \right. .\label{eqav} \qquad\qquad
\end{eqnarray}

\n The first thing to note about eq. (\ref{eqav}) is that the right hand side is an average of four random functions 
whose fluctuations are correlated. The average of the product then involves the product of the averages and other
contributions which come from averages taken in pairs, triplets, and all
four random functions. 

\subsection{Disorder-1induced renormalization of the current terms}

\n At this stage, in order to simplify notation, we shall omit the $L$ index. However, we have to remember that
all terms labelled by indeces $R$ or $\k$ are matrices in $\{L\}$ space, so the order of multiplication of various
terms in the expression has to be
preserved. We shall also omit the $\gamma$ index of the current term indicating the required projection onto a direction.
If required, we can put them back in the final expression.

\n The real-space representation of the random current operator, can take the values ${\bf \j}^{AA}_{RR'}$, 
${\bf \j}^{AB}_{RR'}$, ${\bf \j}^{BA}_{RR'}$, or ${\bf \j}^{BB}_{RR'}$ with probabilities $x_A^2$ , $x_A x_B$, 
$x_B x_A $, and $x_B^2$, respectively. We may rewrite ${\bf \j}_{RR'}$ as
\begin{eqnarray*}
&&\mbf{\j}_{RR'}\eq \mbf{\j}^{AA}_{RR'}\ n_{R}\ n_{R'} + 
\mbf{\j}^{AB}_{RR'}\ n_{R}\ (1-n_{R'})+ \dots \\
&&\qquad \dots \mbf{\j}^{BA}_{RR'}\ (1-n_{R})\ n_{R'}+ \mbf{\j}^{BB}_{RR'}\ (1-n_{R})(1-n_{R'}).
\end{eqnarray*}

\n Following the same augmented-space procedure as for the single-particle Green functions, we get

\begin{widetext}
\begin{eqnarray}
&&{\wt \mbf{\j}}\eq \sum_{R }\sum_{R' }\ \left[\rule{0mm}{5mm}\ \ll \mbf{\j}\gg_{RR' } \ a^{\dagger}_{R }\ a_{R'}+
 (x_B-x_A )\ \mbf{\j}^{(1)}_{RR'}\ a^{\dagger}_{R }\ a_{R' }
\ (b^\dagger_{R }b_{R }\pls b^\dagger_{R' }b_{R' }) \ldots\right. \nonumber\\
&&\qquad \ldots +  \sqrt{x_Ax_B }\; \mbf{\j}^{(1)}_{RR'}\ a^{\dagger}_{R }
\ a_{R' }\ (b_{R }+ b^\dagger_{R }+ b^\dagger_{R' }+ b_{R' }) \ldots\nonumber\\
&&\qquad \ldots + (x_B-x_A )\sqrt{x_A x_B }\; \mbf{\j}^{(2)}_{RR'}\ \ a^{\dagger}_{R }\ a_{R' }
\ \left\{\ b^\dagger_{R }b_{R }\ (b^\dagger_{R' }+b_{R' }) \pls b^\dagger_{R' }b_{R' }\ (b^\dagger_{R }
+b_{R })\right\}\ldots\nonumber\\
&&\qquad \ldots\left. + (x_B-x_A )^2\  \mbf{\j}^{(2)}_{RR'}\ a^{\dagger}_{R }\ a_{R' }
\ b^\dagger_{R' } b_{R' }\ b^\dagger_{R }b_{R }+ x_Ax_B  \ \mbf{\j}^{(2)}_{RR'}\ a^\dagger_{R }\ a_{R' }
\  \left\{ (b^\dagger_{R }+ b_{R })\ (b^\dagger_{R' }+b_{R' })\right\}\rule{0mm}{5mm}\right],
\label{eq12}
\end{eqnarray}
\end{widetext}

\noindent where

\begin{eqnarray*}
 \mbf{\j}^{(1)}_{RR'}&\eq& x_A \ \left( \mbf{\j}^{AA}_{RR'}- \mbf{\j}^{AB}_{RR'}\right)
\mns x_B \left( \mbf{\j}^{BB}_{RR'}\mns \mbf{\j}^{BA}_{RR'}\right), \\
 \mbf{\j}^{(2)}_{RR'}&\eq &\mbf{\j}^{AA}_{RR'}\pls \mbf{\j}^{BB}_{RR'}
\mns \mbf{\j}^{AB}_{RR'}\mns \mbf{\j}^{BA}_{RR'}.
\end{eqnarray*}

\n The first term in Fig.~\ref{fig5} is the averaged current. The figure shows the 15 different scattering 
vertices arising from  terms in eq. (\ref{eq12}). The rule for obtaining the diagrams for the correlation
function $S_\gamma(z_1,z_2)$ is as follows :
Take any two current diagrams from Fig.~\ref{fig5} and two propagators and  join them end to end. 
Now join the configuration fluctuation lines (shown as dashed arrows) in all possible ways.

\begin{figure}[h]
\begin{center}
\framebox{\rotatebox{270}{\includegraphics[width=1.2in, height=3.2in]{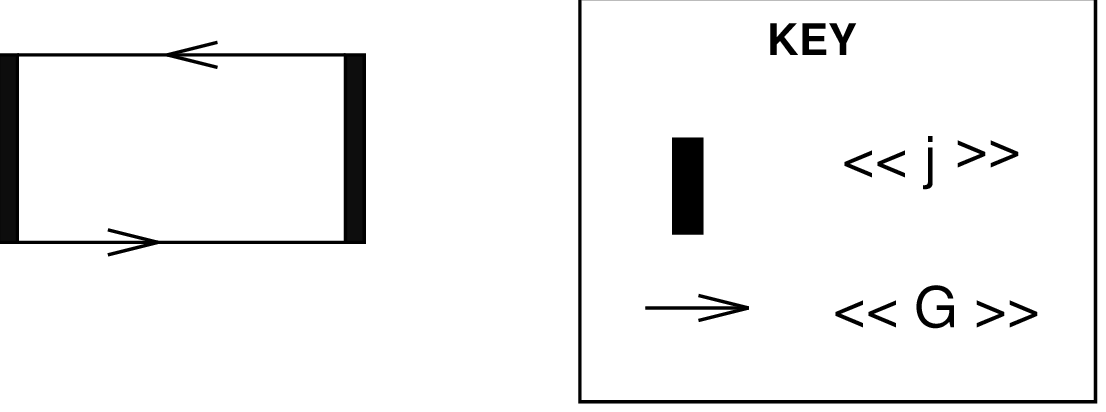}}}
\caption{\label {fig7}
The  diagram for $\left[\ll {\bf j}\gg\ \ll G^v\gg\ \ll {\bf j} \gg\ \ll G^c\gg\right]$}
\end{center}
\end{figure}

\n The dominant contribution comes from the diagram shown in Fig.~\ref{fig7}.
Here the two current terms are the averaged current, and all configuration-fluctuation 
decorations renormalize only the two electron propagators. In this diagram the bold propagators are the fully
renormalized electron propagators and the contribution of this term is 

\begin{widetext}
\be 
\sumk \ll {\mathbf \j}(\k)\gg\ \ll {\mathbf G}^v(\k,z_1)\gg\ \ll {\mathbf \j}(\k)\gg^\dagger\ \ll {\mathbf G}^c(\k,z_2)\gg .
\label{eq16}
\ee
\end{widetext}

\begin{figure}[t]
\begin{center}
\framebox{\includegraphics[width=3.2in, height=4.0in]{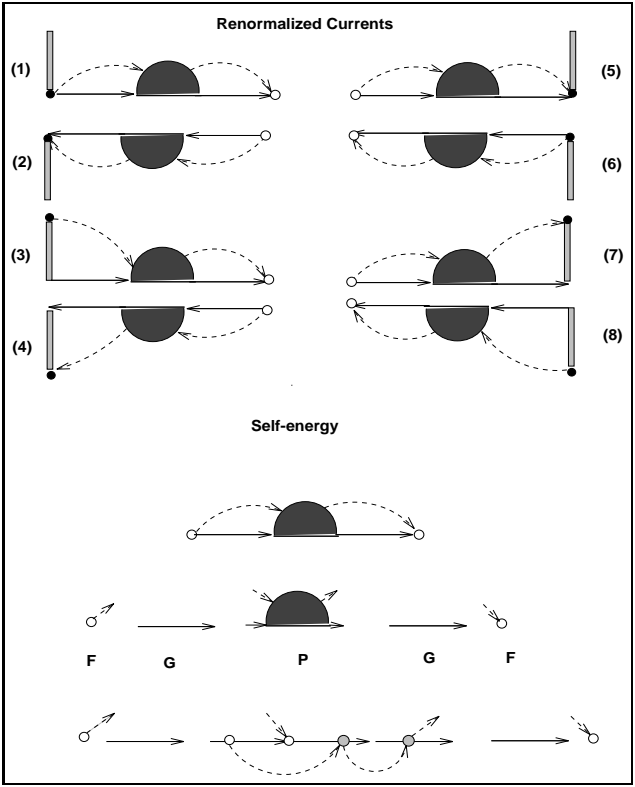}}
\caption{\label {fig8}
Relation between the renormalized currents and the self-energy.}
\end{center}
\end{figure}

We now focus on the main correction terms to the expression in eq.~(\ref{eq16}). These are the correction terms to the
averaged current which are closely related to the self-energies. They arise  from a set
 of diagrams in which no disorder propagator (shown as dashed lines) joins either two
electron propagators or two of the current lines directly.
These diagrams are made out of a left {\sl renormalized} current diagram chosen out of the diagrams (1)-(4) in Fig.~\ref{fig8}
and one right {\sl renormalized} current diagram from (5)-(8)  connected by two {\sl renormalized} propagators : the bottom one being a 
 valence and the top a conduction electron propagator. 

\n Let us now obtain expressions for the renormalized currents. A careful look at the self-energy diagrams 
(see the bottom of Fig.~\ref{fig8} and the example diagram shown there) shows that all 
self-energy diagrams have the structure

\begin{eqnarray}
&&{\mathbf \Sigma}(\k,z)  \eq  {\mathbf F}(z) \ {\mathbf \Phi}(\k,z)\ {\mathbf F}(z), \nonumber\\
&&\phantom{x}\nonumber\\
\mbox{where} && {\mathbf  \Phi}(\k,z) \mbox{\ is the Fourier transform of }\nonumber\\                      
&&\phantom{x}\nonumber\\
&& {\mathbf \Phi}_{RR'}(z) \eq \sum_{R_1R_2} \G_{RR_1}(z)\ P^{RR'}_{R_1R_2}(z)\ \G_{R_2R'}(z).\nonumber
\label{rc1}
\end{eqnarray}

\n While the contribution of the  diagram labelled (1) in Fig.~\ref{fig8} is

\be 
{\mathbf  \j}^{(1)}(\k)\  \widetilde{\mathbf F}(z_1)\ {\mathbf \Phi}(\k,z_1)\  {\mathbf F}(z_1),
\label{rc2}
\ee

\n where

\begin{eqnarray*}
F_{LL'}(z) &\eq& \sqrt{x_Ax_B}\ \frac{1}{f_L(z)}\ \delta_{LL'}, \\
\widetilde{F}_{LL'} &\eq& \sqrt{x_Ax_B} \ \delta_{LL'},
\end{eqnarray*}

\n the expression for (\ref{rc2}) becomes 

\[ {\mathbf \j}^{(1)}(\k)\  \widetilde{\mathbf F}(z_1)\ {\mathbf F}^{-1}(z_1)\ \Sigma(\k,z_1) \eq {\mathbf \j}^{(1)}(\k)\  
{\mathbf f}(z_1)\ {\mathbf \Sigma}(\k,z_1). \]

\n The contributions of  the other diagrams in the left column of Fig.~\ref{fig8} are 

\begin{eqnarray*}
{\mathbf \Sigma}(\k,z_2)\ {\mbf F}^{-1}(z_2)\ \widetilde{\mbf F}(z_2)\ {\mbf \j}^{(1)}(\k) 
&=& {\mbf\Sigma}(\k,z_2)\ \mbf{f}(z_2)\ \mbf{\j}^{(1)}(\k),\\
\widetilde{\mbf F}(z_1)\  {\mathbf \j}^{(1)}(\k)\ {\mbf F}^{-1}(z_1)\ {\mbf \Sigma}(\k,z_1) 
&=& {\mathbf  \j}^{(1)}(\k)\ {\mathbf f}(z_1)\ {\mathbf \Sigma}(\k,z_1),\\
{\mathbf \Sigma}(\k,z_2)\ {\mathbf F}^{-1}(z_2)\  {\mathbf \j}^{(1)}(\k)\ \widetilde{\mbf F}(z_2)
&=& \mbf{\Sigma}(\k,z_2)\ \mbf{f}(z_2)\ \mbf{\j}^{(1)}(\k).
\end{eqnarray*}

\n From the forms of ${\bf F}(z)$ and $\widetilde{\mbf F}(z)$, we note that :

\[ {\mbf \j}^{(1)}(\k)\ \widetilde{\mbf F}(z)\ {\mbf F}(z) = \widetilde{\mbf F}(z)\ {\mbf \j}^{(1)}(\k)\ 
{\mbf F}(z)={\mbf \j}^{(1)}_{LL'}(\k) f_{L'}(z). \]

\n Similarly, the contributions of the diagrams in the right column in Fig.~\ref{fig8} are 

\begin{eqnarray*}
{\mathbf \Sigma}(\k,z_1)\ {\mathbf f}(z_1)\ {\mathbf \j}^{(1)}(\k), && 
{\mathbf \j}^{(1)}(\k)\ {\mathbf f}(z_2)\ {\mathbf \Sigma}(\k,z_2),\\
{\mathbf \Sigma}(\k,z_1)\ {\mathbf f}(z_1)\ {\mathbf \j}^{(1)}(\k), &&  
{\mathbf \j}^{(1)}(\k)\ {\mathbf f}(z_2)\ {\mathbf \Sigma}(\k,z_2).
\end{eqnarray*}

\begin{figure}[h]
\begin{center}
\rotatebox{0}{\includegraphics[width=3.0in, height=1.2in]{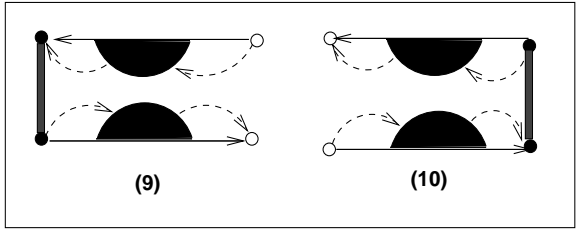}}
\caption{\label {fig9}
Few more renormalized currents.}
\end{center}
\end{figure}

\n Closely related to the above diagrams is a group of diagrams which describe joint fluctuations of one current and two propagators. Two
such  diagrams labelled (9) and (10) in Fig.~\ref{fig9} can also be expressed in terms of the self-energy :

\begin{eqnarray*}
&&\mbf{\Sigma}(\k,z_2)\ \mbf{f}(z_2)\  \mbf{\j}^{(2)}(\k)\ \mbf{f}(z_1)\ \mbf{\Sigma}(\k,z_1), \quad \quad \\
&&\mbf{\Sigma}(\k,z_1)\ \mbf{f}(z_1)\ \mbf{\j}^{(2)}(\k)\  \mbf{f}(z_2)\ \mbf{\Sigma}(\k,z_2).
\end{eqnarray*}

\n If we now gather all the contributions from these diagrams, we may define a renormalized current term as follows~:
\begin{widetext}
\begin{eqnarray}
\mbf{J}^{\rm{eff}}(\k,z_1,z_2)& \eq  \ll {\bf j}(\k)\gg + 2\left[\rule{0mm}{5mm} \mbf{\Sigma}(\k,z_2)\ \mbf{f}(z_2)\ 
\j^{(1)}(\k) \pls \j^{(1)}(\k)\ \mbf{f}(z_1)\ \Sigma(\k,z_1)\right] \ldots\nonumber \\
&\ldots\pls \Sigma(\k,z_2)\ \mbf{f}(z_2)\ \j^{(2)}(\k)\ \mbf{f}(z_1)\ \Sigma(\k,z_1). 
\end{eqnarray}
\end{widetext}

 The  contribution of these disorder renormalized currents and propagators to the correlation function is

\begin{widetext}
\begin{eqnarray}  \ll {\cal S}_{(1)}(z_1,z_2)\gg\ \eq\ \sumk\ \mbox{Tr}\left[\rule{0mm}{4mm}
\mbf{J}^{\rm{eff}}(\k, z_1,z_2)\ll \mbf{G}^v(\k,z_1)\gg
\mbf{J}^{\rm{eff}}(\k,z_1,z_2)^{\dagger}\ll \mbf{G}^c(\k,z_2)\gg\right].\nonumber\\
\end{eqnarray}
\end{widetext}

\begin{figure}[h]
\begin{center}
\framebox{\includegraphics[width=3.2in, height=2.0in]{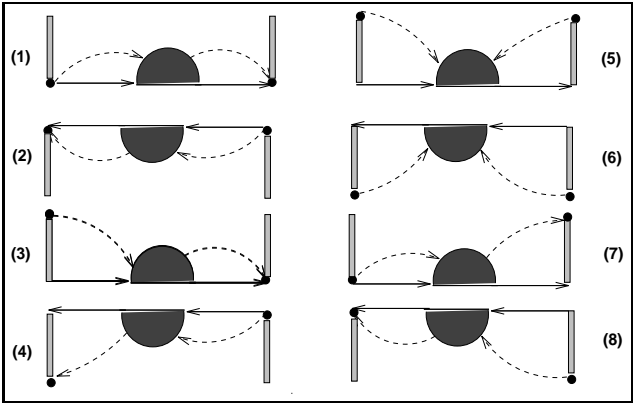}}
\caption{\label {fig11}
The scattering  diagrams associated with joint fluctuations of the random current terms
and one propagator.}
\end{center}
\end{figure}

\n We now turn to terms which involve joint fluctuations between the two current terms and one propagator. We shall show that the corrections due to these terms are also related to the self-energy.
Such diagrams are shown in Fig.~\ref{fig11}. Contribution of these diagrams is given by

\begin{widetext}
\begin{eqnarray}
 \ll {\cal S}_{(2)}(z_1,z_2)\gg  &\eq& 4\ \sumk \ \mbox{Tr} \left [ \ \rule{0mm}{5mm} \j^{(1)}(\k)\ \mbf{f}(z_1)\ \Sigma(\k,z_1)\ \mbf{f}(z_1)\ {\j^{(1)}(\k)}^\dagger \ \ll \mbf{G}(\k,z_2)\gg \right.\ldots\nonumber\\
&& \qquad \ldots\left. \pls {\j^{(1)}(\k)}^\dagger \ \mbf{f}(z_2)\ \Sigma(\k,z_2)\ \mbf{f}(z_2)\ \j^{(1)}(\k)\ \ll\mbf{G}(\k,z_1)\gg \rule{0mm}{5mm}\right].
\end{eqnarray}
\end{widetext}

These terms have a slightly different structure than those shown in Fig.~7. However,
they still depend only on the self-energy.

Intuitively, we expect these to be the dominant disorder scattering correction to the averaged current. It is important to 
note that this correction can be obtained from the self-energy and is therefore eminently computationally feasible in the case of realistic alloys,
once we have a feasible method for obtaining the self-energy. 

There are other scattering diagrams which are not related to the self-energy, but rather
to the vertex corrections. In these diagrams, a disorder line connects both the electron
propagators directly. We expect these corrections to be less dominant. For the sake of completeness,
we shall indicate how to obtain them in the Appendix. We should note that since these
corrections are related to the vertex corrections and we shall indicate how to obtain them
within a ladder approximation, we need not sacrifice these terms in a calculation for a
realistic alloy if we do not wish to do so. However, in most cases we expect their contribution to be relatively small.

\begin{figure*}[t]
\begin{center}
\framebox{\rotatebox{0}{\includegraphics[width=5.0in, height=4.0in]{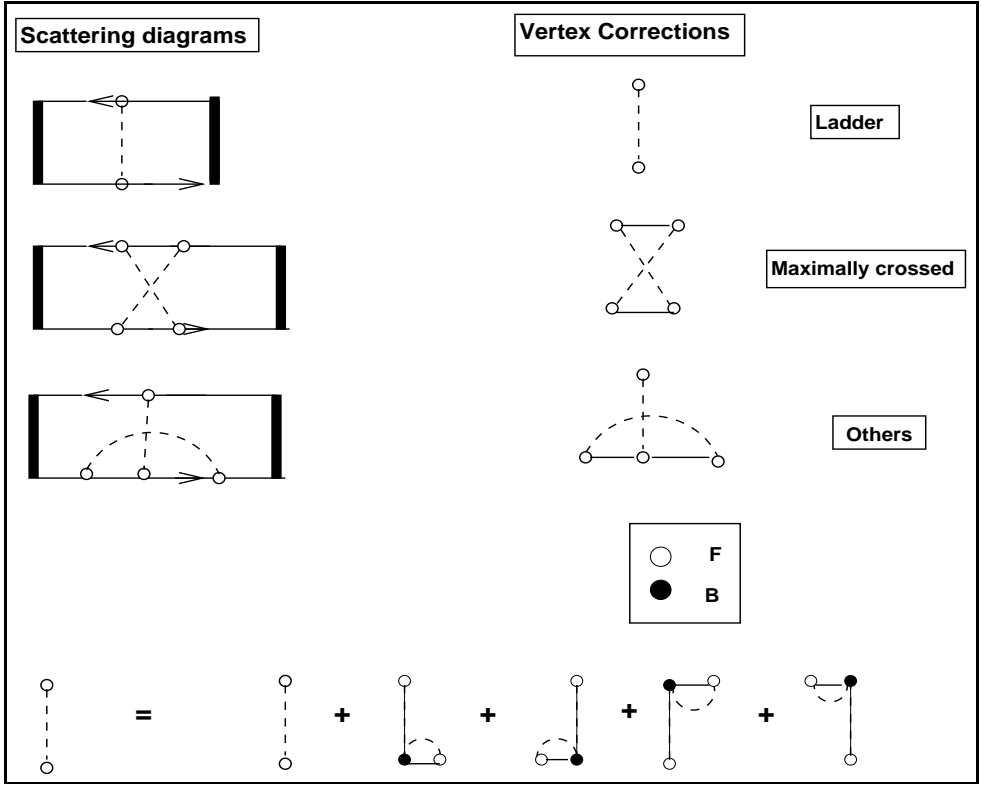}}}
\caption{\label {fig13}
The scattering diagrams leading to vertex corrections.}
\end{center}
\end{figure*}

\subsection{The vertex correction}

We shall now examine the scattering diagrams we have left out, namely those
in which disorder lines connect both the propagators directly. These lead to
vertex corrections due to electron-electron and electron-hole correlated propagation.
Figure~\ref{fig13} show a few of these diagrams. In general, we obtain a Bethe-Salpeter
equation for the averaged two-particle propagator. We shall consider only one special class
of vertex diagrams in this paper, namely the scattering diagrams which are built out of
repeated vertices shown on the first line of Fig.~\ref{fig13}. These are called the
ladder diagrams and can be summed up to all orders. This is the disorder scattering version
of the random-phase approximation (RPA) for electron-electron scattering. There is another
form of diagrams shown on the second line of Fig.~\ref{fig13} with ladder insertions between the 
crossed vertices. These are known as maximally crossed diagrams. These diagrams lead to the
localization effect.

\begin{figure*}[t]
\begin{center}
\framebox{\rotatebox{270}{\includegraphics[width=3.2in, height=5.5in]{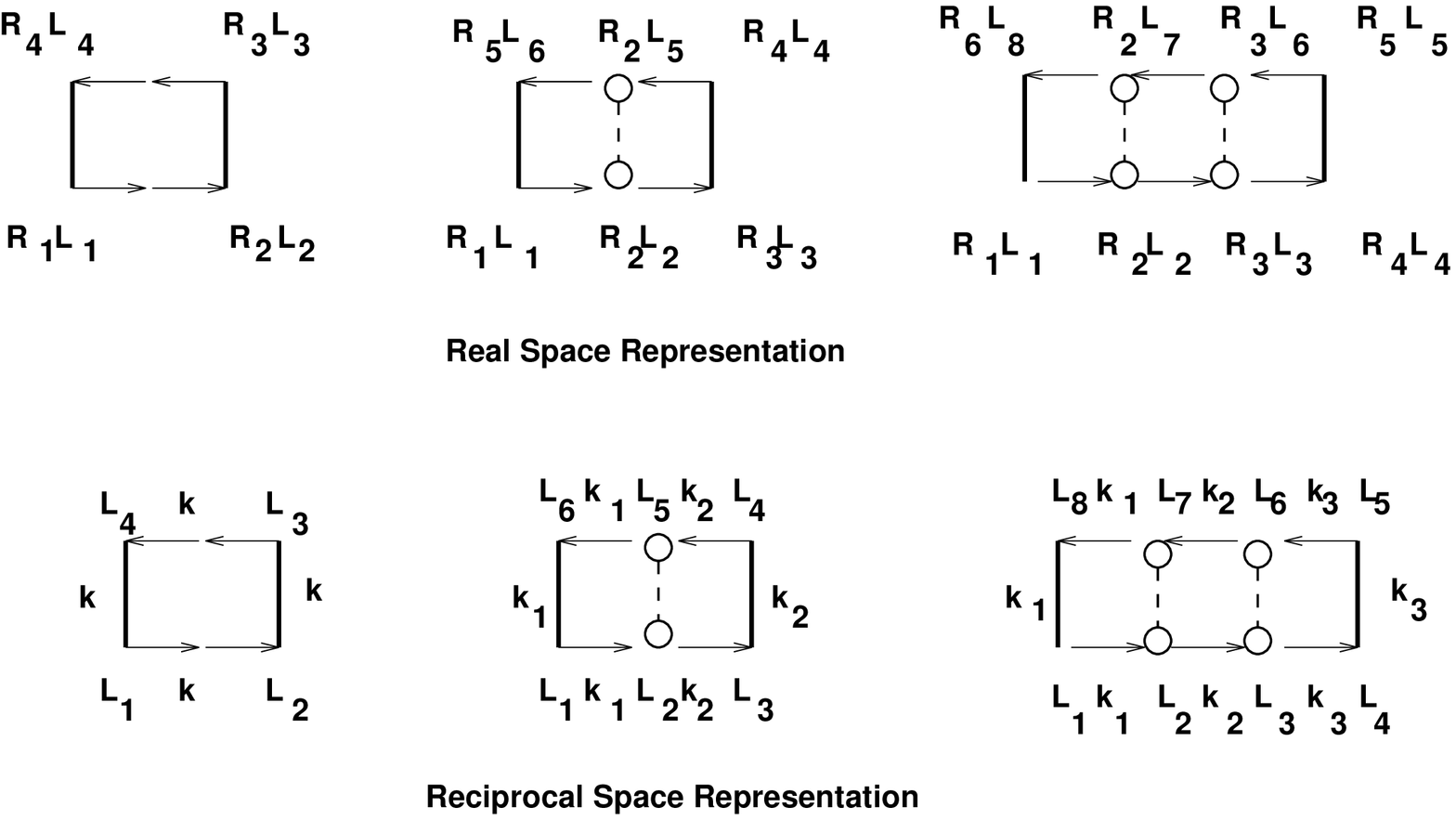}}}
\caption{\label {fig14}
The ladder scattering diagrams for the vertex correction in real-space and reciprocal-space representations.}
\end{center}
\end{figure*}

Here we shall sum the ladder diagrams to all orders. The contribution of the ladder 
diagram shown as the second diagram on the top line in Fig.~\ref{fig14} is 

\begin{widetext}
\begin{eqnarray*}
 \sum_{R_1R_2}\sum_{R_3R_4}\sum_{R_5}\ \sum_{L_1L_2}\sum_{L_3L_4}\sum_{L_5L_6}\ \mbf{J}^{\rm{eff}}_{R_5L_6,R_1L_1}
\ G_{R_1L_1,R_2L_2}(z_1) \ W^{L_5}_{L_2}\  G_{R_2L_2,R_3L_3}(z_1)\ldots\\
\phantom{xxxxxxxxx} \ldots \mbf{J}^{\rm{eff}}_{R_3L_3,R_4L_4}\!\!\!\!\!\!\!\!\phantom{i}^\dagger\ \ G_{R_4L_4,R_2L_5}(z_2)
 \ G_{R_2L_5,R_5L_6}(z_2)
\end{eqnarray*}
\end{widetext}

\n where $G_{RLR'L'}(z)\ =\ \ll G^{v/c}_{RLR'L'}(z)\gg$  and 

\begin{eqnarray*}
W^L_{L'} = F_L(z_2) \left[ \delta_{LL'}+2 \sum_{L^{''}} \left[\rule{0mm}{4mm} B_{L^{''}}(z_1) \ G_{RL^{''},RL'}(z_1) \right.\right.\dots \\
\left.\left.\dots \ +B_{L^{''}}(z_2)\  G_{RL^{''},RL'}(z_2)\right]\right]F_{L'}(z_1). \qquad
\end{eqnarray*}

Homogeneity in augmented-space means that this is independent of $R$ and it allows us to take the 
Fourier transforms leading to

\begin{eqnarray}
\left[ \sumk\ \mbf{G}(\k,z_2)\ \mbf{J}^{\rm{eff}}(\k,z_1,z_2)\ \mbf{G}(\k,z_1)\right]\dots\qquad \nonumber \\ 
\ldots\ \mbf{W} \ \left[\int_{\rm{BZ}}\frac{d^3\k'}{8\pi^3}\ \mbf{G}(\k',z_1)
\ \mbf{J}^{\rm{eff}}\!\!\phantom{i}^{\dagger}(\k',z_1,z_2)\ \mbf{G}(\k',z_2)\right]\nonumber\\
\eq  \mbf{\Gamma}(z_1,z_2)\enskip \mbf{W}\enskip \widehat{\mbf{\Gamma}}(z_1,z_2)\qquad
\label{str}
\end{eqnarray}

\n We define

\begin{eqnarray*}
\int_{\rm{BZ}} \frac{d^3\k}{8\pi^3}\ \mbf{G}(\k,z_2)\ \mbf{J}^{\rm{eff}}(\k,z_1,z_2)\ \mbf{G}
(\k,z_1) = \mbf{\Gamma}(z_1,z_2) \qquad\\
\int_{\rm{BZ}} \frac{d^3\k'}{8\pi^3}\ \mbf{G}(\k',z_1)\ 
\mbf{J}^{\rm{eff}}(\k',z_1,z_2)\!\!\phantom{i}^{\dagger} \mbf{G}(\k',z_2)=\widehat{\mbf{\Gamma}}(z_1,z_2). \quad
\end{eqnarray*}

Let us now look at the contribution of the set of ladder diagrams. Each one of them has the same structure
as eq.~(\ref{str}). We may then sum up the series as follows.

\n Let us define
\begin{eqnarray*}
&&\lambda^{L_1L_2}_{L_3L_4}(z_1,z_2) \eq \int_{\rm{BZ}}\ \frac{d^3\k}{8\pi^3}\ 
G_{L_3L_4}(\k,z_1)\ G_{L_2L_1}(\k,z_2), \\
&&\omega^{L_1L_2}_{L_3L_4} \eq  W^{L_1}_{L_3}\ \delta_{L_1L_2}\ \delta_{L_3L_4}.
\end{eqnarray*}

\n These super-matrices in $\{L\}$ space are written as $\uu\lambda$ and $\uu\omega$. The full ladder vertex may now be written as

\begin{eqnarray}
\uu\Lambda(z_1,z_2) &=&\uu\omega+ \uu\omega\ \uu\lambda(z_1,z_2)\ \uu\omega + \uu\omega\ \uu\lambda(z_1,z_2)
\ \uu\omega\ \uu\lambda(z_1,z_2)\ \uu\omega + \ldots\nonumber\\
 &=& \uu\omega \left( \uu I \mns \uu\lambda(z_1,z_2)\ \uu\omega \right)^{-1}
\end{eqnarray}

\n The ladder diagram vertex correction now can be written as 

\begin{widetext}
\begin{eqnarray}
\ll S_{\rm{ladder}}(z_1,z_2)\gg \ =\ \mbox{Tr}
\sum_{L_1L_2}\sum_{L_3L_4}\Gamma^{L_1}_{L_2}(z_1,z_2)\ \Lambda^{L_1L_3}_{L_2L_4}\ \widehat{\Gamma}^{L_3}_{L_4}(z_1,z_2)
=\ \mbox{Tr} \ \mbf{\Gamma}(z_1,z_2)\otimes \mbf{\widehat{\Gamma}}(z_1,z_2)\ {\Lambda}(z_1,z_2). \quad\qquad
\end{eqnarray}
\end{widetext}

\section{Comments and Conclusion}

Starting from the pseudo-fermion picture in the augmented space method, we have obtained an expression for the configuration-averaged 
optical conductivity. The disorder scattering renormalizes both the electron propagators as well as the current terms. We have shown 
that the dominant corrections to the averaged current can be related to the self-energy. For the sake of completeness, we have also 
shown that the remaining correction terms are related to the vertex corrections. We have also indicated how to obtain the vertex 
corrections within the ladder approximation.  Once we set up a computationally feasible technique for the computation of the self-energy 
and the ladder approximation to the vertex correction, all the correction terms can be easily obtained. In an earlier communication 
\cite{kasr}, we have suggested the augmented-space recursion as a feasible technique for obtaining $\Sigma(\bf k)$ and have applied it for
obtaining the complex band structure and density of states of a series of realistic metallic alloys, namely AgPd and AuFe  and most recently 
NiPt among others. We propose to use that technique and the results derived here to obtain the configuration-averaged optical conductivity 
in disordered metallic alloys. We intend to study, through numerical calculations,
the relative importance of the contribution of the different correction terms. 

\begin{appendix} 

\section{Corrections to the current term related to the vertex corrections}

\begin{figure*}[t]
\begin{center}
\framebox{\includegraphics[width=3.0in, height=1.8in]{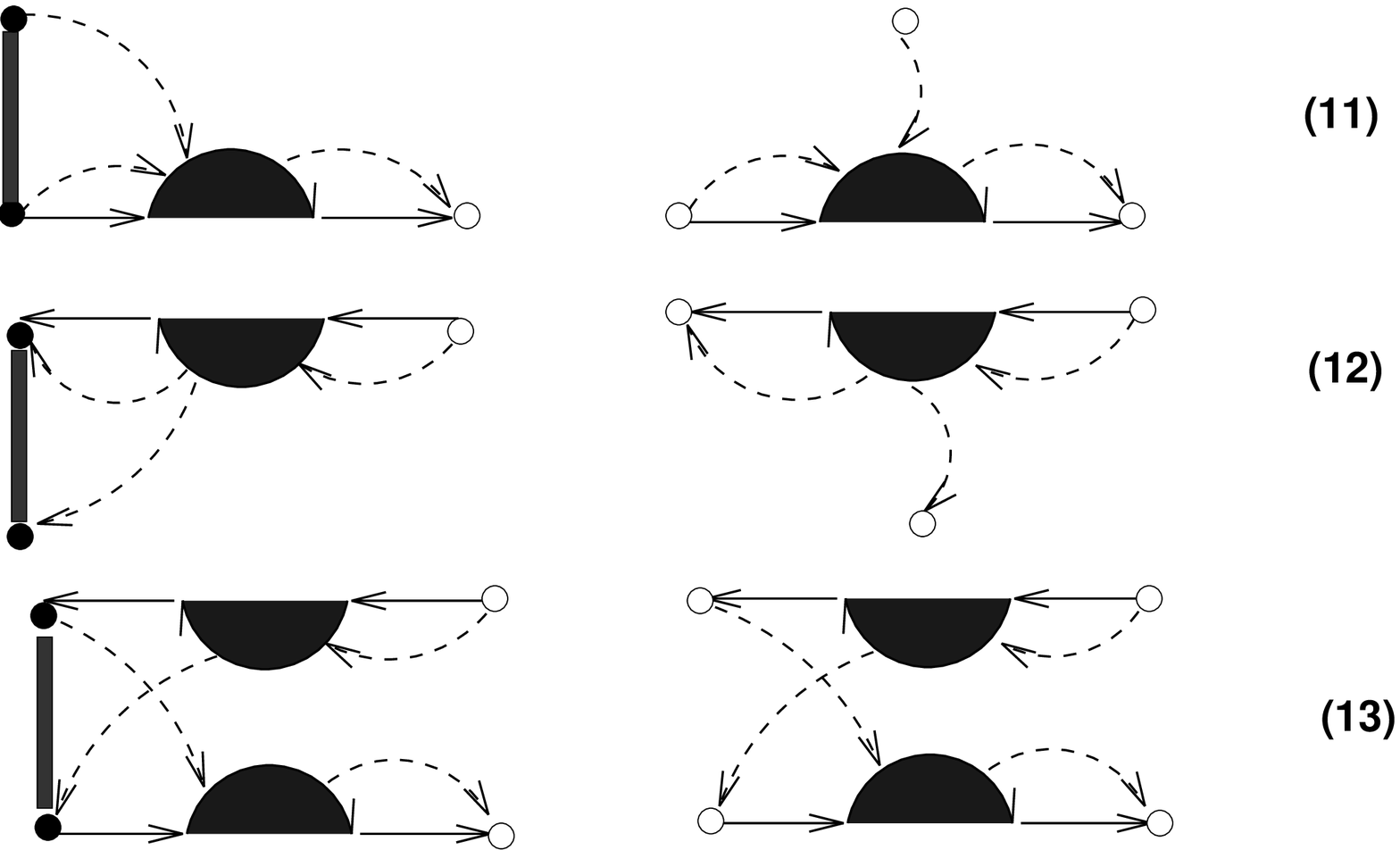}}
\framebox{\includegraphics[width=3.0in, height=1.8in]{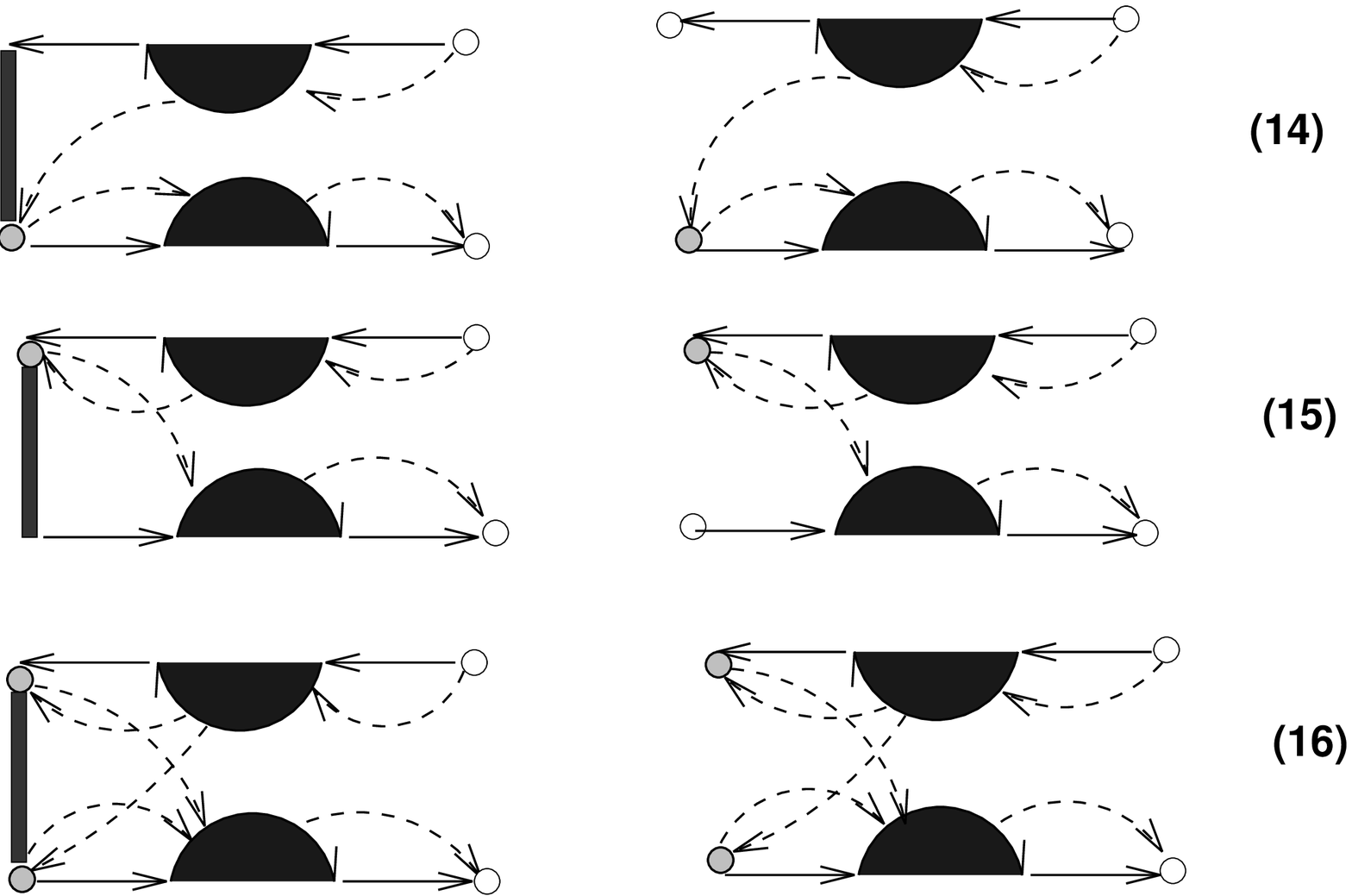}}
\vspace{0.2cm}\phantom{x}\vskip 0.2cm
\framebox{\includegraphics[width=3.0in, height=1.4in]{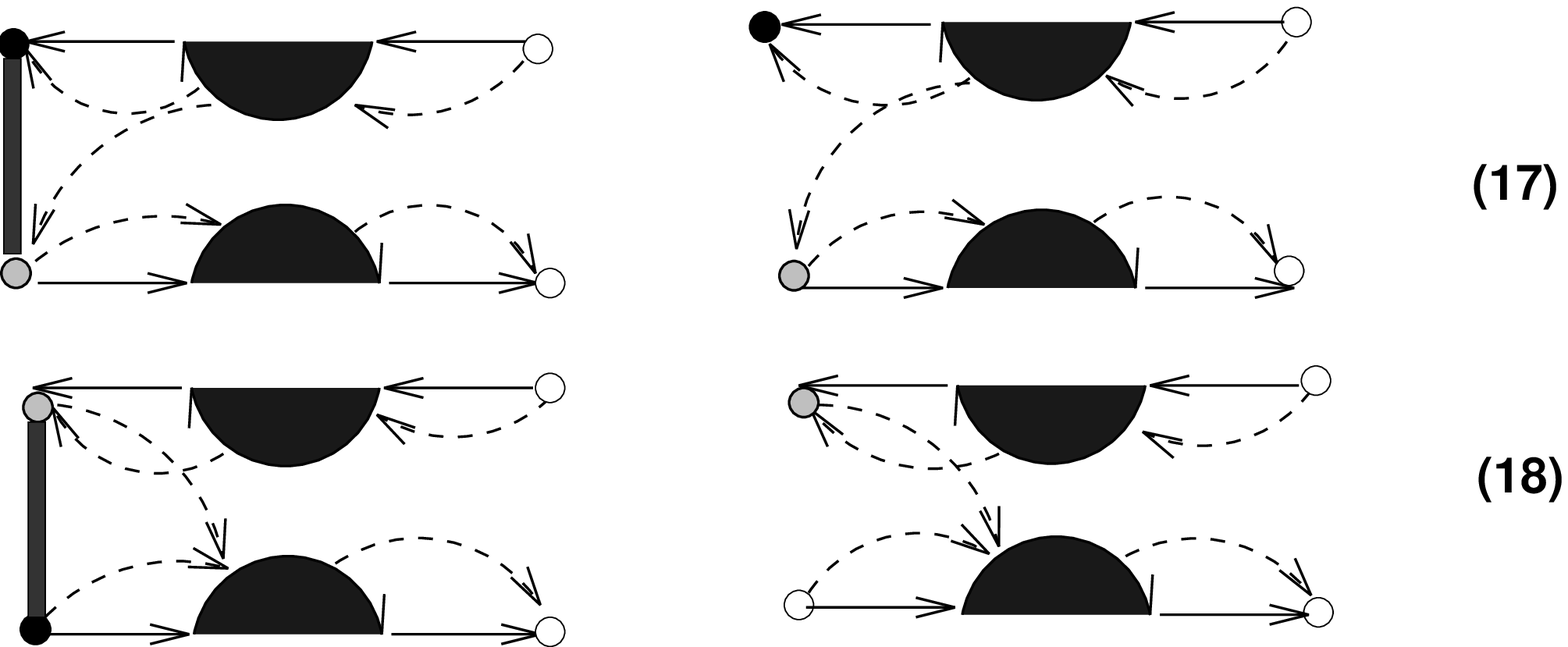}}
\vspace{0.2cm}\phantom{x}\vskip 0.2cm
\caption{\label {fig10}
Renormalized currents (left column) derived from vertex corrections (right column).}
\end{center}
\end{figure*}

For the sake of completeness, we shall also indicate the contribution of
those  scattering diagrams to the current which cannot be directly related to the
self-energy but rather
 to vertex corrections.

 These remaining diagrams are shown in the left column of Fig.~\ref{fig10}. These diagrams cannot be
related to the self-energies, but rather to specific vertex correction diagrams between the two propagators. There are
three categories of diagrams~: ones that involve $\widetilde{\mbf F}$ vertices [labelled (11)-(13)], ones that involve 
$\widetilde{\mbf B}\ $[labelled (14)-(16)] type vertices and those that involve both [labelled (17)-(18)].

The first category of diagrams (11)-(13) contributes the following~:

\begin{eqnarray*}
\widetilde{\mbf F}(z_1)\ {\mbf F}^{-1}(z_1)\ {\mbf \j}^{(2)}(\k)\  \widetilde{\mbf F}(z_2)\ {\mbf F}^{-1}(z_2)\ {\Lambda}^{(F)}({\mbf 0},\k;z_1,z_2)\\
\quad\Rightarrow\quad \mbox{diagram (11)},\\
 \widetilde{\mbf F}(z_1)\ {\mbf F}^{-1}(z_1)\ {\mbf \j}^{(2)}(\k)\  \widetilde{\mbf F}(z_2)\ {\mbf F}^{-1}(z_2)\ {\Lambda}^{(F)}(\k,{\mbf 0};z_1,z_2)\\
\quad\Rightarrow\quad \mbox{diagram (12)},\\
\widetilde{\mbf F}(z_1)\ {\mbf F}^{-1}(z_1)\ {\mbf \j}^{(2)}(\k)\  \widetilde{\mbf F}(z_2)\ {\mbf F}^{-1}(z_2)\ {\Lambda}^{(F)}(\k,\k;z_1,z_2)\\
\quad\Rightarrow\quad \mbox{diagram (13)}.
\end{eqnarray*}

\n Inserting the expressions for ${\bf F}(z)$ and $\widetilde{\bf F}$, we get a total contribution,

\begin{eqnarray} 
&&{\mbf J}_1\ =\ {\mbf f}(z_1)\ {\mbf \j}^{(2)}(\k)\ {\mbf f}(z_2)\left\{\rule{0mm}{5mm}
{\Lambda}^{(F)} ({\mbf 0},\k,z_1,z_2) \right.\dots \nonumber\\
&&\quad \left.\dots + {\Lambda}^{(F)} (\k,{\mbf 0},z_1,z_2) +{\Lambda}^{(F)} (\k,\k,z_2,z_1)\right\}.
\end{eqnarray}

\n Here, the vertex correction term ${\Lambda}^{(F)}$ involves only $F$-like vertices in all four legs.
Similarly, for the other two sets of diagrams we get

\begin{eqnarray*}
&&\widetilde{\mbf B}(z_1)\ {\mbf B}^{-1}(z_1)\ {\mbf \j}^{(1)}(\k)\ {\mbf F}^{-1}(z_2)\ {\Lambda}^{(B)}(\k,\k;z_1,z_2)\\
&&\phantom{xxxxxxxxxxxxxxxxxxxxxxxx} \Rightarrow \quad \mbox{diagram (14)},\\
&&{\mbf F}^{-1}(z_1)\ {\mbf \j}^{(2)}(\k)\ \widetilde{\mbf B}(z_2)\ {\mbf B}^{-1}(z_2)\ {\Lambda}^{(B)}(\k,\k;z_1,z_2) \\
&&\phantom{xxxxxxxxxxxxxxxxxxxxxxxx} \Rightarrow \quad \mbox{diagram (15)},\\
&&\widetilde{\mbf B}(z_1)\ {\mbf B}^{-1}(z_1)\ {\mbf \j}^{(2)}(\k)\ \widetilde{\mbf B}(z_2)\ {\mbf B}^{-1}(z_2)\ {\Lambda}^{(B)}(\k,\k;z_1,z_2) \\
&&\phantom{xxxxxxxxxxxxxxxxxxxxxxxx} \Rightarrow \quad \mbox{diagram (16)}.
\end{eqnarray*}

\n The total contribution will be 

\begin{eqnarray}
&&{\mbf J}_2\ =\ \left\{\rule{0mm}{5mm} \frac{2}{\sqrt{x_Ax_B}}\ \f(z_1)\ {\mbf \j}^{(1)}(\k)\ \f(z_2)\ +\ \right.\dots\nonumber\\ 
&&\quad \left.\dots \f(z_1)\ {\mbf \j}^{(2)}(\k)\ \f(z_2)\right\} {\Lambda}^{(B)}(\k,\k;z_1,z_2).
\end{eqnarray}

${\Lambda}^{(B)}$ involves only $B$-like vertices in its left-hand side legs.  Finally for the last two diagrams, 

\begin{eqnarray*}
\widetilde{\mbf B}(z_1)\ {\mbf B}^{-1}(z_1)\ {\mbf \j}^{(2)}(\k)\ \widetilde{\mbf F}(z_2)\ {\mbf F}^{-1}(z_2)\ {\Lambda}^{(FB)}(\k,\k;z_1,z_2) \\
\quad\quad\Rightarrow\quad \mbox{diagram (17)}.\\
\widetilde{\mbf F}(z_1)\ {\mbf F}^{-1}(z_1)\ {\mbf \j}^{(2)}(\k)\ \widetilde{\mbf B}(z_2)\ {\mbf B}^{-1}(z_2)\ {\Lambda}^{(FB)}(\k,\k;z_1,z_2) \\
\quad\quad\Rightarrow\quad \mbox{diagram (18)}.\\
\end{eqnarray*}

\n Their contribution is

\begin{equation}
{\mbf J}_3\ =\ 2\ \f(z_1)\ {\mbf \j}^{(3)}(\k)\ \f(z_2)\ {\Lambda}^{(FB)}(\k,\k;z_1,z_2).
\end{equation}

\n Collecting together terms 

\[ \mbf{\Delta J}  =  \mbf{J}_1 + \mbf{J}_2 + \mbf{J}_3. \]

\begin{figure}[b]
\begin{center}
\framebox{\includegraphics[width=3.4in, height=1.8in]{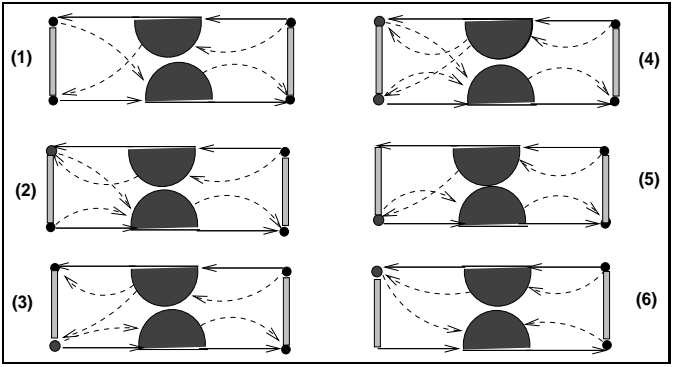}}
\caption{\label {fig12}
Some of the scattering  diagrams associated with joint fluctuations of the random current terms
and two propagators.}
\end{center}
\end{figure}

 The contribution of these disorder renormalized currents and propagators to the correlation function is 

\begin{widetext}
\begin{eqnarray}  \ll {\cal S}_{(3)}(z_1,z_2)\gg\ &\eq\ &\sumk\ \mbox{Tr}\left[\rule{0mm}{4mm}
\mbf{\Delta J}(\k, z_1,z_2)\ll \mbf{G}^v(\k,z_1)\gg
\mbf{\Delta J}(\k,z_1,z_2)^{\dagger}\ll G^c(\k,z_2)\gg\right].
\end{eqnarray}
\end{widetext}

\n Finally, Fig.~\ref{fig12} shows the diagrams with joint fluctuations of two current terms and two propagators.
These are also built out of vertex corrections. 
Note that each of the six diagrams can be broken up
into a left and right part. For the diagrams shown in Fig.~\ref{fig12} all the right parts are the same.
Thirty other similar diagrams can be produced by replacing the right part with the five different left
parts mirror-imaged. The contribution of these diagrams is then, if

\begin{widetext}
\[
\mbf{K}(\k,z_1,z_2)\eq  \mbf{f}(z_2)\ \j^{(2)}(\k)\ \mbf{f}(z_1)+ \mbf{b}(z_2)\ \j^{(2)}(\k)\ \mbf{f}(z_1) 
 +\mbf{f}(z_2)\ \j^{(2)}(\k)\ \mbf{b}(z_1)+ \mbf{b}(z_2)\ \j^{(2)}(\k)\ \mbf{b}(z_1),
\]
\end{widetext}

\n  where

\[ \mbf{b}(z) \eq \frac{x_B-x_A}{\sqrt{x_Ax_B}}\ \mbf{f}(z), \]

\begin{widetext}
\be
\ll {\cal S}_{(4)}(z_1,z_2)\gg  
\eq \sumk \ \mbox{Tr}\ \mbf{K}(\k,z_1,z_2)\otimes\mbf{K}^\dagger(\k,z_1,z_2)\ {\Lambda}(\k,\k,z_1,z_2) .
\ee
\end{widetext}
\end{appendix}


\newpage
\begin{thebibliography}{99}
\bibitem{Ander1} Andersen O. K., \prb {\bf 12} 3060 (1975)
\bibitem{Ander2} Jepsen O. and Andersen O. K., 1971 {\sl Solid St. Commun.} {\bf 9} 1763 
\bibitem{Am} Mookerjee A., 1973 {\sl J. Phys.: Condens. Matter} {\bf 6} 1340 
\bibitem{KG} Kaplan T. and Gray L.J., \prb {15} 3260 (1977) 
\bibitem{Am2} Saha T., Dasgupta I. and Mookerjee A., 1996 {\em J. Phys.: Condens. Matter} {\bf 8} 1979 
\bibitem{Am4} Dasgupta I., Saha T. and Mookerjee A., 1997 {\em J. Phys.: Condens. Matter} {\bf 9} 3529 
\bibitem{Am3} Ghosh S., Das N. and Mookerjee A., 1999 {\it Int. J. Mod. Phys.} { \bf B 21} 723    
\bibitem{ziman} Ziman J.M., 1961 {\it Phil. Mag.} {\bf 6} 1013
\bibitem{bradley} Bradley C.C., Faber T.E., Wilson E.G. and Ziman J.M., 1962 {\it Phil. Mag.} {\bf 7} 865
\bibitem{fz} Faber T.E. and Ziman J.M., 1965 {\it Phil. Mag.} {\bf 11} 153
\bibitem{evans} Evans R., Greenwood D.A. and Lloyd P., 1971 {\it Phys. Lett.} {\bf 35A} 57
\bibitem{rs} Roth L. and Singh V.A., 1980 {\it J. Physique} {\bf C8} 459
\bibitem{roth} Roth L., \prb {\bf 9} 2476 (1974)
\bibitem{rs1} Singh V.A. and Roth L., 1980 {\it Bull. Am. Phys. Soc.} {\bf 25} 242
\bibitem{ay} Asano S. and Yonezawa F., 1980 {\em J. Phys. F: Met. Phys.} {\bf 10} 75
\bibitem{v} Velick\'y B., 1969 {\em Phys. Rev.} {\bf 184} 614
\bibitem{bv} Brouers F. and Vedyayev A.V., \prb {\bf 5} 348 (1972)
\bibitem{nii} Niizeki K., 1977 {\em J. Phys. C: Solid State Phys.} {\bf 10} 211 ; {\bf 10} 2131 ; {\bf 10} 2141
\bibitem{hw} Hoshino K. and Watabe M., 1977 {\it J. Phys. Soc. Japan} {\bf 43} 583
\bibitem{nh} Niizeki K. and Hoshino K., 1977 {\em J. Phys. C: Solid State Phys.} {\bf 10} 3351
\bibitem{mook2} Mookerjee A., 1976 {\em J. Phys. C: Solid State Phys.} {\bf 9} 1225
\bibitem{leath} Leath P.L., \prb {\bf 2} 3078 (1970)
\bibitem{hp} Harris R. and Plischke M., 1972 {\em Solid St. Commun.} {\bf 11} 1165
\bibitem{nr} Nauciel-Bloch M. and Riedinger R., 1974 {\em J. Phys. F: Met. Phys.} {\bf 4} 1032
\bibitem{mook3} Mookerjee A., 1975 {\em J. Phys. C: Solid State Phys.} {\bf 8} 1524
\bibitem{mty} Mookerjee A., Thakur P.K. and Yussouff M., 1985 {\em J. Phys. C: Solid State Phys.} {\bf 18} 4677
\bibitem{mt} Mookerjee A. and Thakur P.K., 1988 {\em J. Phys. C: Solid State Phys.} {\bf 21} 943
\bibitem{shultz} Schultz T.D. and Shapero D., 1973 {\em Phys. Rev.} {\bf 181} 1062
\bibitem{gdma} Ghosh S., Das N., Mookerjee A., Huda A., Ahmed A. and Halder A.,1990 {\sl Int. J. Mod. Phys} B 
\bibitem{kn:asf} Kumar V., Mookerjee A. and Srivastava V.K., 1982, {\em J. Phys. C: Solid State Phys.} {\bf 15} 1939
\bibitem{hhk} Haydock R., Heine V. and Kelly M.J., 1972, {\em J. Phys. C: Solid State Phys.} {\bf 5} 2845
\bibitem{vol35} Haydock R. {\sl Solid State Physics} vol 35 (Academic Press, New York)
\bibitem {kn:hay} Haydock R., 1980, {\sl Solid State Phys.} {\bf 35} 216; 1981, {\sl Phil. Mag.} {\bf B 43} 203 ; Haydock R. and Te. R. L., \prb {\bf 49} 10845 (1994)
\bibitem {kn:hay2} Haydock R., {\sl thesis} University of
 Cambridge, 1972
\bibitem{bp} Beer N. and Pettifor D. G. (1982), in {\sl The Electronic Structure of Complex Systems},
ed. P. Phariseau and W. M. Temmerman, NATO ASI Series B, v.113, p 769
\bibitem{kn:sdm} Saha T., Dasgupta I and Mookerjee A., 1994, {\sl J. Phys.: Condens. Matter} {\bf 6} L245
\bibitem{kasr} Saha K.K. and Mookerjee A., Cond-Mat/0405175 (2004) ; Biswas P., Sanyal B., Mookerjee A., Huda A., Chowdhury N., Ahmed M. and Halder A., {\it Int. J. Mod. Phys.} {\bf B11} 3703 ; Biswas P., Sanyal B., Fakhruddin M., Halder A., Mookerjee A. and Ahmed M., 
{\it J. Phys. Condens Matter} {\bf 7} 8569 (1995) 
\end{thebibliography}
\end{document}